\documentstyle[12pt,amsfonts]{article}
\parskip=\medskipamount

\def\appendix#1{
  \addtocounter{section}{1}
  \setcounter{equation}{0}
  \renewcommand{\thesection}{\Alph{section}}
 \section*{Appendix \thesection\protect\indent
 \parbox[t]{11.715cm} {#1}}
 \addcontentsline{toc}{section}{Appendix \thesection\ \ \ #1}
  }
\renewcommand{\thefootnote}{\fnsymbol{footnote}}

\newcommand {\cD}{{\cal D}}
\newcommand {\cE}{{\cal E}}

\newcommand {\cG}{{\cal G}}

\newcommand {\cN}{{\cal N}}

\newcommand {\cP}{{\cal P}}

\newcommand{\bB}{{\bf B}}

\newcommand{\bE}{{\bf E}}

\newcommand{\bH}{{\bf H}}

\newcommand{\bW}{{\bf W}}

\def\a{\alpha}
\def \bi{\bibitem}

\def\b{\beta}

\def\d{\delta}
\def\e{\epsilon}

\def\g{\gamma}

\def\j{\psi}
\def\k{\kappa}
\def\l{\lambda}
\def\m{\mu}

\def\o{\omega}

\def\q{\theta}
\def\r{\rho}
\def\s{\sigma}

\def\D{\Delta}
\def\F{\Phi}

\def\O{\Omega}

\def\S{\Sigma}

\newcommand{\ad}{{\dot{\alpha}}}                           
\newcommand{\bd}{{\dot{\beta}}}                            

\newcommand{\pa}{\partial}                           
\newcommand{\hf}{\frac12}

\newcommand{\vf}{\varphi}
\newcommand{\be}{\begin{equation}}
\newcommand{\ee}{\end{equation}}
\newcommand{\bea}{\begin{eqnarray}}
\newcommand{\eea}{\end{eqnarray}}
\newcommand{\non}{\nonumber}
\begin{document}

\begin{titlepage}
\thispagestyle{empty}

\begin{flushright}
hep-th/0109183 \\
September, 2001
\end{flushright}
\vspace{5mm}

\begin{center}
{\Large \bf Goldstone multiplet for partially broken \\
superconformal symmetry}
\end{center}
\vspace{3mm}

\begin{center}
{\large
S.M. Kuzenko and I.N. McArthur
}\\
\vspace{2mm}

${}$\footnotesize{
{\it
Department of Physics, The University of Western Australia\\
Crawley, W.A. 6009. Australia}
} \\
{\tt  kuzenko@cyllene.uwa.edu.au},~
{\tt mcarthur@physics.uwa.edu.au}
\vspace{2mm}

\end{center}
\vspace{5mm}

\begin{abstract}
\baselineskip=14pt
The bosonic parts of D3-brane actions in AdS(5) backgrounds
are known to have symmetries which are field-dependent extensions 
of conformal transformations of the worldvolume coordinates. 
Using the coset space $SU(2,2|1)/SO(4,1)$, 
we apply the method of nonlinear realizations to construct 
a four-dimensional $\cN=1$ off-shell supersymmetric action 
which has a generalized field-dependent superconformal invariance. 
The Goldstone fields for broken scale, chiral and 
S-supersymmetry transformations form a chiral supermultiplet.  
\end{abstract}

\vfill
\end{titlepage}

\newpage
\setcounter{page}{1}

\renewcommand{\thefootnote}{\arabic{footnote}}
\setcounter{footnote}{0}

In the consideration of D3-branes in an $AdS_5 \times S^5$ background, 
Maldacena \cite{M} showed that the  scale-invariant action
\be
S = -\kappa^2 \int {\rm d}^4x \, U^4 \left( \sqrt{ - {\rm det} (\eta_{m
n} +  \partial_{m}U \partial_{n} U/U^4)} - 1
\right)
\label{radial}
\ee
is also invariant under the generalized special conformal
transformations
\begin{eqnarray}
\delta x^m &=& b^m \,x^2 -2x^m (b \cdot x)
+ b^m/ U^2 (x)~,
\nonumber \\
\delta U(x) & \equiv & U'(x') - U(x) = 2 (b \cdot x)\,
U(x)~.
\label{deform}
\end{eqnarray}
This action and its symmetries have a natural geometric origin. The action
is the Nambu action, expressed in static
gauge, for a 3-brane embedded in  $AdS_5$ with the metric
\be
{\rm d} s^2 = U^2 {\rm d} x^m \eta_{mn} {\rm d}x^n + ({\rm d}U/U)^2~.
\ee
The  symmetries of the action follow
from the $SO(4,2)$ group of isometries of $AdS_5 = SO(4,2)/SO(4,1)$.

The action (\ref{radial}) captures the dynamics of the transverse 
radial excitation of a D3-brane  moving near the core of another 
D3-brane. The bosonic D3-brane action reads
 (here we omit the Chern-Simons term; see, for example, \cite{T}
for more details),
\be
S = - \kappa^2 \int {\rm d}^4x \, U^4 \left( \sqrt{ - {\rm det} (\eta_{m
n} +  \partial_{m}U^\m \partial_{n} U^\m/U^4
+ F_{mn} / U^2) } - 1~
\right) ~,
\label{d3brane}
\ee
where $U^2 = U^\m U^\m$, with $\m= 1,\ldots, 6$. 
It is invariant under standard linear conformal 
transformations of $U$ and $F$
only if all terms in the action with derivatives 
of the transverse brane excitations, $\pa_m U^\m$, are ignored. 
In the general case, when such terms are kept,
the action turns out to be invariant under deformed 
nonlinear conformal transformations of the fields
which have a  deep five-dimensional origin
(see \cite{Claus,Metsaev} and references therein)
but which, at first sight, look wierd from a four-dimensional 
point of view. Remarkably, there does exist a four-dimensional 
origin for this deformed conformal symmetry, 
and in fact it comes from quantum theory. 
Almost two decades  ago, Fradkin and Palchik \cite{FP} observed that 
the effective action 
in conformally invariant quantum Yang-Mills theories
{\it is not}  invariant under the usual 
linear conformal transformations of non-abelian gauge
fields. Instead, it possesses an invariance under 
nonlocal and nonlinear transformations consisting of 
a combination of  linear conformal transformations
and some compensating field-dependent gauge transformations, 
which form a nonlinear realization of the conformal group \cite{Pal}
(see also \cite{FP2}). 
These nonlocal conformal transformations turn out to reduce to 
those of the form ({\ref{deform}) in the case of the low-energy 
effective action of $\cN=4$ super Yang-Mills theory \cite{JKY}.

In the present letter, we carry out the first steps 
towards solving the problem 
of constructing off-shell supersymmetric {\it four-dimensional} 
extensions of the bosonic action (\ref{d3brane}). 
An $\cN$-extended superfield generalization of (\ref{d3brane})
should be invariant under nonlinearly realized 
$SU (2,2|\cN )$ transformations. Such nonlinear superfield realizations
of $SU(2,2|\cN)$ could in principle be obtained by 
(i) considering an off-shell $\cN=1$ or $\cN=2$ 
formulation of the quantum $\cN=4$ super Yang-Mills theory;
(ii) defining quantum nonlocal superconformal transformations
of the dynamical  superfields \`{a} la Fradkin and Palchik
\cite{FP}; (iii) considering a low-energy approximation defined
in a similar manner to the component approach of 
\cite{JKY} but now in terms
of superfields. Once the nonlinear superfield realization of 
$SU(2,2|\cN)$ is known, an invariant action should be 
determined to a significant extent  on general grounds \cite{M,T}. 
In practice, the above  program may be difficult to realize. 
It seems simpler and safer to make use of the method of nonlinear 
realizations \cite{nlr}, in particular as applied to the conformal 
symmetry \cite{ISS}, 
 in order to  work out consistent mechanisms 
for partial breaking of the superconformal symmetry.
The coset spaces of interest are of the form   
$SU(2,2|\cN)/ (SO(4,1) \times G_R)$, where $G_R$ is a subgroup 
of the $R$-symmetry group $U(\cN)$. 

In this paper, we give an $\cN=1$ supersymmetric extension 
of the action 
\be
S = - \kappa^2 \int {\rm d}^4x \, U^4 \left( \sqrt{ - {\rm det} (\eta_{m
n} +  \partial_{m}U^\m \partial_{n} U^\m/U^4 ) } - 1
\right) ~, \qquad \m =1,2~, 
\ee
which corresponds\footnote{The $\cN=1$ superconformal 
Born-Infeld action given in eq. (2.15) of \cite{KT2} reduces
in components to the action (\ref{d3brane}) with 
$\m=1,2$ and all derivatives of $U$ omitted.}
 to the choice
$F=0$ and $\m=1,2$ in (\ref{d3brane}).
The  action is constructed in terms of a chiral scalar 
superfield which is the Goldstone multiplet 
for partial breaking of $SU(2,2|1)$.
We expect that it models the essential features of 
nonlinearly realized superconformal symmetries 
of the low-energy $\cN=4$ super Yang-Mills effective 
action and D3-brane actions. 

The supergroup $SU(2,2|1)$ is the four-dimensional 
$\cN=1 $ superconformal group. It   
is generated by Lie algebra elements
of the form\footnote{We use the superconformal notation adopted
in \cite{O,KT1}. Our two-component matrix-like conventions are as follows. 
${}$For all $Q$-supersymmetry spinors, namely  transformation
parameters $( \e, \bar \e) $ and  superspace coordinates
$(\q, \bar \q )$, the matrix convention is: $\j =(\j^\a),~ 
\tilde{\j}  = (\j_\a),~ {\bar \j}= ({\bar \j}^\ad ),~
 \tilde{\bar \j} = ({\bar \j}_\ad)$.
${}$For all $S$-supersymmetry spinors, the matrix convention is:
 $\j =(\j_\a),~ 
\tilde{\j}  = (\j^\a),~ {\bar \j}= ({\bar \j}_\ad ),~
 \tilde{\bar \j} = ({\bar \j}^\ad)$. 
${}$For a four-vector $x^a$ converted into a bi-spinor, 
 we define $x= ( x_{\a \ad} )$ and $\tilde{x} = (x^{\ad \a})$;
 but $x \cdot y = x^a y_a = - \hf {\rm tr} (\tilde{x} y )
 = - \hf {\rm tr} (x \tilde{y})$.}
\be
X = \left(
\begin{array}{ccc}
\o_\a{}^\b - \D \d_\a{}^\b  \quad &  -{\rm i} b_{\a \bd} \quad &
2\r_\a \\
 -{\rm i} a^{\ad \b} \quad & -{\bar \o}^\ad{}_\bd 
+ {\bar \D}  \d^\ad{}_\bd   \quad &
2{\bar \e}^{\ad}  \\
2\e^\b \quad & 2{\bar \r}_{ \bd} \quad & 2({\bar \D} - \D)
\end{array}
\right)
\label{algebra}
\ee
which satisfy the conditions
\be
{\rm str} \;X = 0~, \qquad
B X^\dag B = - X~, \qquad 
B = \left( 
\begin{array}{ccc}
0 \quad & {\bf 1} \quad & 0 \\
{\bf 1} \quad & 0 \quad & 0 \\
0 \quad & 0 \quad & -1 
\end{array}
\right)~.
\ee
As superconformal transformations, the matrix elements correspond to a
Lorentz transformation $(\o_\a{}^\b,~{\bar \o}^\ad{}_\bd)$,
a translation $a^{\ad \a}$, a special conformal transformation
$ b_{\a \ad}$, a $Q$--supersymmetry $(\e^\a,~ {\bar \e}^{\ad })$,
an $S$--supersymmetry $(\r_\a,~{\bar \r}_{ \ad})$,
and a combined scale and chiral transformation 
$\D = \hf (\l  - \frac{\rm i}{3} \O )  $. 
Minkowski superspace, with coordinates
$z^A = (x^a, \q^\a ,{\bar \q}_\ad )$,  
 can be identified with the coset space
$SU(2,2|1) / (\cP \times {\Bbb C}^* )$, 
where $\cP$ denotes the $\cN=1$ 
Poincar\'e supergroup generated by the parameters 
$(\o_,~{\bar \o}, ~ b,~ \r,~ {\bar \r} )$ in (\ref{algebra}), 
and ${\Bbb C}^*$ denotes the group of scale and chiral transformations
generated by the parameters $\D$ and $\bar \D$ in (\ref{algebra}).
The coset representative $g(z)$ has a matrix realization
\be
g (z) = 
\left(
\begin{array}{ccc}
{\bf 1} \quad &      0  \quad  &      0  \\
-{\rm i} \tilde{x}_+ \quad& {\bf 1} \quad & 2{\bar \q} \\
2\q  \quad & 0 \quad & 1 
\end{array} \right)
\ee
where $x_\mp $ denote ordinary (anti-)chiral bosonic variables,
$x^a_\pm = x^a \pm {\rm i} \q \s^a {\bar \q}.$
In the infinitesimal case,
the superconformal group acts on Minkowski superspace 
by transformations $z \rightarrow z' = z + \d z$
such that   
\be
 X\; g (z)= \d g (z) + g (z)\; H(z)~, 
\qquad \d g(z) = g(z + \d z ) - g(z)~,
\ee
where the matrix
\be
H (z) = \left(
\begin{array}{ccc} 
\hat{\o}
- \s {\bf 1}
\quad &  -{\rm i} b
\quad &
2 \hat{\r}{} \\
 0 \quad & - \hat{{\bar \o}}
+  {\bar \s}  {\bf 1}
\quad &
0  \\
0  \quad & 2 \hat{ {\bar \r}} \quad & 
{2}( {\bar \s} - \s  )
\end{array}
\right)
\label{stability}
\ee
is a compensating transformation 
belonging to the Lie algebra of the 
stability group.  One finds 
\bea
\d x^a_+ &=& v^a (x_+, \q) = a^a + \o^a{}_b x^b_+
+ \l x^a_+  +b^a x^2_+  - 2 x^a_+ (x_+ \cdot b) 
+ 2 {\rm i} \q \s^a {\bar \e} - 2 \q \s^a \tilde{x}_+ \r ~, \non \\
\d \q^\a &=& v^\a (x_+ , \q) =
\e^\a - \q^\b \o_\b{}^\a + \hf (\l + {\rm i} \O ) \q^\a
+ (\q b \tilde{x}_+)^\a - {\rm i} ({\bar \r} \tilde{x}_+)^\a
+ 2 \r^\a \q^2 ~,
\label{hol}
\eea
and the entries of $H$ read 
\bea
\hat{\o}_\a{}^\b (x_+, \q) &=& \o_\a{}^\b  
+ \hf (x_+ \tilde{b} - b \tilde{x}_+)_\a{}^\b + 4 \r_\a \q^\b
+ 2 \d_\a{}^\b \q \r~, \non \\
\s (x_+, \q) &=& \D + 2 \q \r - b\cdot x_+ ~, 
\qquad 
\hat{\r}_\ad (\q ) = 2( {\bar \r}  + {\rm i} \q b)_\ad
=  {\bar D}_\ad {\bar \s}~,
\eea  
with $D_\a$ and ${\bar D}_\ad$ the flat spinor covariant derivatives.
As is seen from (\ref{hol}), the superconformal group 
acts on the complex variables $x_+ $ and $\q$ by holomorphic 
transformations which preserve the surface $x_+^a - x_-^a =
2{\rm i} \q \s^a {\bar \q}$; see \cite{BK} for a detailed
discussion. 
The left-invariant Maurer-Cartan one-form 
\be
E = g^{-1} \,{\rm d} g =
\left(
\begin{array}{ccc}
0 ~  &    0  ~  &      0  \\
-{\rm i} \tilde{e} ~ & 0 ~ & 2{\rm d} {\bar \q} \\
2{\rm d} \q  ~ & 0 ~& 1 
\end{array} \right) ~, \qquad
e^a = {\rm d} x^a - {\rm i} {\rm d} \q \s^a {\bar \q} 
+ {\rm i}  \q \s^a {\rm d} {\bar \q}~,
\ee
transforms only under the compensating transformations,
\be 
\d E = [H, E ] - {\rm d} H~.
\ee

We now turn to the consideration of the coset 
space $SU(2,2|1) / SO(4,1) $, which describes
the product of  a five-dimensional $\cN=1$ anti-de Sitter superspace
and a one-sphere, $AdS^{5|8} \times S^1$.
This space can be parametrized in terms of a coset   
representative $G(Z)$ with the  matrix realization
\bea
G (Z) &=& 
\left(
\begin{array}{ccc}
{\bf 1} ~ &      0  ~  &      0  \\
-{\rm i} \tilde{x}_+  ~& {\bf 1} ~ & 2{\bar \q} \\
2\q  ~ & 0 ~ & 1 
\end{array} \right)
~
\left(
\begin{array}{ccc}
{\bf 1} ~  &    2 \eta {\bar \eta}   ~  &      2 \eta  \\
0 ~ & {\bf 1} ~ & 0 \\
0  ~ & 2 {\bar \eta}  ~& 1 
\end{array} \right) ~
\left(
\begin{array}{ccc}
\vf ^{1/2} {\bf 1} ~  &   0   ~  &    0 \\
0 ~ & {\bar \vf}^{-1/2} {\bf 1} ~ & 0 \\
0  ~ & 0   ~& \vf / {\bar \vf} 
\end{array} \right) 
\non \\
&=& \left(
\begin{array}{ccc}
\vf^{1/2} {\bf 1} \quad &
2{\bar \vf}^{-1/2} \eta {\bar \eta}
& 2 {\vf}{\bar \vf}^{-1}  \eta \\
 -{\rm i} \vf^{1/2} \tilde{x}_+ \quad &
{\bar \vf}^{-1/2} \big(1- 2{\rm i} \tilde{x}_+  \eta {\bar \eta}
 +4 {\bar \q} {\bar \eta} \big) {\bf 1}
\quad &
2 {\vf}{\bar \vf}^{-1}
\big( {\bar \q}  - {\rm i} \tilde{x}_+ \eta \big) \\
2\vf^{1/2} \q \quad & 2{\bar \vf}^{-1/2} \big( \bar \eta
+ 2 \q \eta {\bar \eta}  \big) \quad &
{\vf}{\bar \vf}^{-1}   \big( 1 + 4 \q \eta \big)
\end{array}
\right) ~.
\eea
An infinitesimal superconformal transformation 
$Z \rightarrow Z' = Z + \d Z$ of the 
variables $Z = (z^A, \vf, {\bar \vf}, \eta_\a , {\bar \eta}_\ad)$
is uniquely determined by requiring 
\be
X \,G (Z) = \d G(Z) + G(Z) \,{\bf H}(Z), \qquad 
\d G(Z) = G(Z+ \d Z ) - G(Z)~, 
\ee
with ${\bf H} $ a compensating $SO(4,1)$ transformation 
of the form (see also \cite{CRZ})
\be
\bH  = \left(
\begin{array}{ccc} 
{\bf \O}
~ &  -{\rm i} {\bf b} 
~ & 0 \\
{\rm i} \tilde{{\bf b}}  ~ & - {\bf {\bar \O}}
~ & 0  \\
0  ~ & 0 ~ & 0
\end{array}
\right)~, \qquad {\rm tr} \,{\bf \O} =0~, \quad 
 {\bf {\bar \O}} = {\bf \O}^\dag ~, \quad {\bf b}^\dag = {\bf b}.
\ee
One finds 
\bea
\d x_+^{\ad \a} &=& v^{\ad \a} (x_+, \q) 
+ (\vf {\bar \vf } )^{-1/2} \Big( {\bf b}^{\ad \a}
+ 4 {\bar \q}^{\ad} {\bar \eta}_{\bd} {\bf b}^{\bd \a} \Big) ~, \non \\
\d \q^\a & = & v^\a (x_+ , \q) - {\rm i} (\vf {\bar \vf} )^{-1/2}
{\bar \eta}_{\bd} {\bf b}^{\bd \a} ~;
\label{TR1} \\
\d  {\rm ln}  \vf & = & - 2 \s (x_+ , \q ) +2{\rm i} 
( \vf {\bar \vf} )^{-1/2} {\bar \eta} \tilde{ \bf b } \eta~, \non \\
\d \eta_\a &=& \hf D_\a \s +(\s - 2{\bar \s}) \eta_\a 
+\hat{\o}_\a{}^\b \eta_\b 
- 4{\rm i} (\vf {\bar \vf} )^{-1/2} \eta_\a 
({\bar \eta} \tilde{\bf b} \eta) ~, 
\label{TR2}
\eea
and the entries of $\bf H$ are
\bea
 (\vf {\bar \vf})^{1/2}  {\bf b}  &=& b 
(1 + (\vf {\bar \vf})^{-1/2} \eta^2 {\bar \eta}^2 ) 
+ 2{\rm i} ( \r {\bar \eta} -\eta {\bar \r})
+2 (b {\bar \q} {\bar \eta}
+ \eta \q b)~, \non \\
{\bf \O} &=& \hat{\o} (x_+, \q) 
+ 2{\rm i} (\eta {\bar \eta} \tilde{\bf b} 
+\hf  {\bar \eta} \tilde{\bf b} \eta \, {\bf 1} )~. 
\eea
As is seen from eq. (\ref{TR1}), 
the variations $\d x^a_+$ and $\d \q^\a$ satisfy the relation
$$
\d x^a_+ - \d x^a_- = 2{\rm i} \,
(\d \q \s^a {\bar \q} + \q \s^a \d {\bar \q} )~,
$$
in complete agreement with the definition of $x_+$. 
However, the variations $\d x_+$ and $\d \q$
are no longer holomorphic functions of  $x_+$ and $\q$, 
unlike the Minkowski superspace case.

The standard coset construction using the Maurer-Cartan form
 $G^{-1}\, {\rm d} G$ yields the geometry $ AdS^{5|8} \times S^1$.
 The coefficients of the
``broken'' generators of $SU(2,2|1)$ determine a supervielbein
$\bE$, and the
coefficients of the ``unbroken'' generators are the
components of an $SO(4,1)$ connection $\bf \S$. 
In terms of the coordinates introduced above,
\bea
G^{-1}\, {\rm d} G &=& \bE + {\bf \S} 
=
\left(
\begin{array}{ccc}
\hf {\bf d} \ln \vf {\bf 1} ~ &
-\frac{\rm i}{2}  {\bf d} x ~
& 2 {\bf d}  \eta \\
 -\frac{\rm i}{2} {\bf d} \tilde{x} ~ &
-\hf {\bf d}\ln{\bar \vf} {\bf 1}
~ & 2  {\bf d} {\bar \q}  \\
2 {\bf d} \q ~ & 2 {\bf d}{\bar \eta}
~ & {\bf d} \ln (\vf / {\bar \vf} ) 
\end{array}
\right)
+ 
\left(
\begin{array}{ccc}
\bW ~ &  -\frac{\rm i}{2}  \bB ~ & 0 \\
 \frac{\rm i}{2} \tilde{\bB} ~ & -{\bar \bW} ~ & 0\\
0 ~ & 0 ~ & 0
\end{array}
\right)~.
\eea
The components of the supervielbein are given 
by scale and chiral invariant one-forms 
\bea
{\bf d} \q & = & \vf ^{-1/2} \,{\bar \vf} \,
\Big( {\rm d} \q +{\rm i} {\bar \eta} \tilde{e} \Big)~,
 \\
{\bf d}  \eta  &=&   \vf^{1/2} \, {\bar \vf}^{-1} \,
\Big( {\rm d} \eta + 2 \eta (
{\bar \eta} {\rm d} {\bar \q } - 2 {\rm d} \q \eta
-{\rm i} {\bar \eta} \tilde{e} \eta ) \Big)~, \\
{\bf d} \ln \vf &=& {\rm d} \ln \vf + 4 {\rm d} \q \eta
+ 2{\rm i} {\bar \eta} \tilde{e} \eta ~, \\
{\bf d} x &=& (\vf \bar \vf )^{1/2} \,e
+ 2 {\rm i}\,(\vf \bar \vf )^{-1/2}
\Big( {\rm d} \eta {\bar \eta} - \eta {\rm d} \bar \eta
+4 \eta \big( {\bar \eta} {\rm d} {\bar \q }
-{\rm d} \q \eta
+\frac{\rm i}{2} {\bar \eta} \tilde{e} \eta \big)
\bar \eta \Big) ~,
\eea
and the components of the $SO(4,1)$ connection  read
\bea
{\bf W} &=& -4 \Big( \eta {\rm d} \q
+ \hf {\rm d} \q \, \eta \,{\bf 1} \Big)
-2{\rm i} \Big(\eta {\bar \eta} \tilde{e}
+ \hf {\bar \eta} \tilde{e} \eta \,{\bf 1} \Big)~, \\
{\bf B}  &=& -(\vf \bar \vf )^{1/2} \,e
+ 2 {\rm i}\,(\vf \bar \vf )^{-1/2}
\Big( {\rm d} \eta {\bar \eta} - \eta {\rm d} \bar \eta
+4 \eta \big( {\bar \eta} {\rm d} {\bar \q }
-{\rm d} \q \eta
+\frac{\rm i}{2} {\bar \eta} \tilde{e} \eta \big)
\bar \eta \Big) ~.
\eea

The superconformal transformations of $\bE$ and $\bf \S$ are 
\be
\d \bE = [ \bH , \bE ] ~, \qquad 
\d {\bf \S} = [ \bH , {\bf \S} ] - {\rm d} \bH~.
\ee 
Analysing the transformation of $\bE$, one observes that 
the supermetric
\bea
{\rm d} s^2 &=& {\bf d}x^a {\bf d} x_a +
\big( {\bf d} \ln \sqrt{ \vf {\bar \vf} } \big)^2 
- \zeta ({\bf d} \ln  \vf / {\bar \vf}  )^2 \non \\ 
&+& \kappa \big( {\bf d} \q^\a {\bf d} \q_\a
-{\bf d} {\bar \eta}_\ad {\bf d} {\bar \eta}^\ad \big)
+{\bar \kappa} \big(
{\bf d} {\bar \q}_\ad {\bf d} {\bar \q}^\ad
- {\bf d} \eta^\a {\bf d} \eta_\a \big) ~,
\label{susymetric}
\eea
with $\zeta$ and $\kappa$ dimensionless constants, 
is superconformally invariant.

So far, $\vf 
$ and $\eta_\a$ have been considered as
independent coordinates
of $AdS^{5|8} \times S^1$.
${}$From now on,
we will treat them as Goldstone superfields, $\vf (z) $ and
$\eta_\a (z) $, living in the four-dimensional $\cN=1$ 
superspace with coordinates $z^M = (x^m, \q^\m, {\bar \q}_{\dot{\m}})$.
On this space, we introduce the supervielbein\footnote{This
supervielbein is used as an $SO(3,1)$ covariant basis 
of one-forms in $\cN=1$ superspace. However, 
the four-dimensional geometry which will be used
to construct a superconformally invariant action will be 
that inherited from the supermetric (\ref{susymetric}).}
\be
\cE^A = ({\bf d} x^a , {\bf d} \q^\a, {\bf d} {\bar \q}_\ad)
= {\rm d} z^M \, \cE_M{}^A (z) ~,
\ee
and the dual basis in the space of vector fields,
\be
\cD_A = (\cD_a, \cD_\a, {\bar \cD}^\ad ) = \cE_A{}^M (z) \;
\frac{\pa}{\pa z^M}~, \qquad \quad
\cE_A{}^M \, \cE_M{}^B = \d_A{}^B~.
\ee
Explicitly, the operators $\cD_A$ take the form
\bea
\left(
\begin{array}{c}
-\hf \cD_{\a \ad} \\
\cD_\a \\
-{\bar \cD}_\ad 
\end{array}
\right)
=
\left(
\begin{array}{ccc}
(A^{-1})_{\a \ad}{}^{\bd \b} ~ &
-{\rm i} (A^{-1})_{\a \ad}{}^{\bd \b} {\bar \eta}_\bd ~ &
{\rm i} (A^{-1})_{\a \ad}{}^{\bd \b} \eta_\b  
\\
(B)_{\a }{}^{\bd \b} ~ &
\frac{ \sqrt{\vf} }{\bar \vf} \d_\a{}^\b 
-{\rm i} (B)_{\a }{}^{\bd \b} {\bar \eta}_\bd  ~ & 
{\rm i} (B)_{\a }{}^{\bd \b} \eta_\b
 \\
(C)_\ad{}^{\bd \b} ~ & 
-{\rm i} (C)_\ad{}^{\bd \b} {\bar \eta}_\bd  ~&
\frac{ \sqrt{ \bar \vf } }{\vf} \d_\ad{}^\bd
+ {\rm i} (C)_\ad{}^{\bd \b} \eta_\b  
\end{array}
\right)
{}
\left(
\begin{array}{c}
-\hf \pa_{\b \bd} \\
D_\b \\
-{\bar D}_\bd 
\end{array}
\right) ~,
\non
\eea
where 
\bea
(\vf {\bar \vf})A_{\a \ad}{}^{\bd \b} &=& 
(\vf {\bar \vf})^{3/2}
\Big(1 + 3 (\vf {\bar \vf})^{-1} \eta^2 {\bar \eta}^2 \Big)
\d_\a{}^\b \d_\ad{}^\bd 
+{\rm i} \Big( \eta^\a (\pa_{\a \ad} {\bar \eta}^\bd) - 
(\pa_{\a \ad} \eta^\a )  {\bar \eta}^\bd \Big) \non \\
&&+ {\bar \eta}^2 (D_\a \eta^\b) \d_\ad{}^\bd 
-\eta^2 ({\bar D}_\ad {\bar \eta}^\bd ) \d_\a{}^\b 
+ 2(D_\a {\bar \eta}^\bd ){\bar \eta}_\ad \eta^\b
+2 ({\bar D}_\ad \eta^\b) \eta_\a {\bar \eta}^\bd ~, \non \\
B_{\a}{}^{\bd \b} &=& -2 {\rm i} {\bar \vf}^{-3/2} 
\Big(  (D_\a \eta^\g ) {\bar \eta}^{ \dot{\g} }  
+ \eta^\g ( D_\a {\bar \eta}^{\dot{\g}}  )
+ 2 \eta^2 {\bar \eta}^{\dot{\g}} \d_\a{}^\g \Big)
(A^{-1})_{ \g \dot{\g} }{}^{\b \bd} ~, \non \\
 C_{\ad}{}^{\bd \b} &=& ~2 {\rm i} { \vf}^{-3/2} 
\Big(  ({\bar D}_\ad \eta^\g ) {\bar \eta}^{ \dot{\g} }  
+ \eta^\g ( {\bar D}_\ad {\bar \eta}^{\dot{\g}}  )
- 2 \eta^\g {\bar \eta}^2 \d_\ad{}^{\dot{\g}} \Big)
(A^{-1})_{ \g \dot{\g} }{}^{\b \bd} ~.
\eea

The Goldstone superfields must be constrained, since 
$\vf$ and $\eta$ contain not only true Goldstone fields
for the broken scale, chiral and $S$-supersymmetry 
transformations, but in addition a number of ghost fields.
Guided by the inverse Higgs effect \cite{IO},
which has played an important role in the construction 
of various models for partial supersymmetry breaking \cite{BG,BIK}, 
a covariant set of constraints is obtained by setting to zero
the coefficients of $\cE_{\a}$
and $\bar{\cE}^{\ad}$ in the decomposition of ${\bf d} \ln \vf$ with
respect to the basis $\cE^A.$
The resulting constraints read:
\bea
0 & = &  \cD_\a \, \ln \vf 
+4 \vf^{1/2} {\bar \vf}^{-1} \eta_\a
+ 2 (\vf \bar{\vf} )^{-1} 
\left( \eta^2 \cD_\a \bar{\eta}^2 - \bar{\eta}^2 
\cD_\a \eta^2 \right)~, \non \\
0 &=& {\bar \cD}_\ad \, \ln \vf + 2 (\vf \bar{\vf} )^{-1} \left(
\eta^2  {\bar \cD}_\ad \bar{\eta}^2 - \bar{\eta}^2 
{\bar \cD}_\ad \eta^2 \right) ~.
\label{constraints}
\eea
These constraints are invariant under the action of the group
$SU(2,2|1),$ because the compensating $SO(4,1)$ transformations
mix the one-form $ {\bf d} \ln \vf $ with ${\bf d} x^a,$ but not
with ${\bf d} \theta^\a$ or ${\bf d} \bar{\theta}_\ad.$
The constraints can be solved by expressing $\vf$ and $\eta$
in terms of a {\it chiral unconstrained} superfield $\F$, 
${\bar D}_\ad \F=0$. To lowest orders in perturbation theory, 
we have
\bea
\ln \vf &=& \ln \F -\frac{1}{ 64} (\F {\bar \F})^{-1}
{\bar D}_\ad \Big[ ({\bar D}^\ad \ln {\bar \F} )
(D \ln \F )^2 \Big]  + O(\ln^4 \F)~, \\ 
\eta_\a &=&  -\frac{1}{4} \, 
 D_\a \ln \vf -\frac{\rm i}{ 64} (\F {\bar \F})^{-1}  \\
& \times &
 \Big[ (D_\a D^\b \ln \F ) ({\bar D}^\bd \ln {\bar \F}) 
+ (D^\b \ln \F ) (D_\a {\bar D}^\bd \ln {\bar \F}) \Big]
\pa_{\b \bd} \ln \F  
 + O (\ln^4 \F) ~ . \non 
\eea

Treating $\vf$ and $\eta$ as Goldstone fields allows
us to consider the pull back of the supermetric (\ref{susymetric})
to four-dimensional superspace,
\be
{\rm d}s^2 = {\rm d} z^M \, \cG_{MN} (z)\, {\rm d}z^N~.
\ee
Then, the action 
\be 
S = |\k|^2  \int {\rm d}^4 x {\rm d}^4 \q \,
\sqrt{- {\rm sdet} (\cG_{MN}) }
\label{ac1}
\ee
is invariant under the nonlinearly realized superconformal 
transformations (\ref{TR1}), (\ref{TR2}). 
This is the Goldstone multiplet action. 
To lowest order in powers of derivatives of the Goldstone chiral 
superfield $\F$, the action reads
\be 
S = \int {\rm d}^4 x {\rm d}^4 \q \,{\bar \F} \F +  O(\ln^4 \F)~.
\ee
The form of this action is not completely specified by 
the requirement of superconformal symmetry, 
as the supermetric (\ref{susymetric}) contains the free 
parameters $\k$ and $\zeta$. This is in constrast to the
bosonic action (1). 
A more detailed analysis of the explicit structure of the action 
(\ref{ac1}) will be given in a separate publication.

It would be of interest to find a closed solution of
the constraints (\ref{constraints}) in terms of $\F$ and its 
conjugate. The approach of \cite{RT} may be useful for this purpose, 
but it would first be necessary to study the geometry of 
the superspace $SU(2,2|1)/SO(4,1)$ along the lines
of \cite{IS}. 
In principle, one could also use the coset space 
$SU(2,2|1)/ \big( SO(4,1) \times U(1) \big)$ to
develop a mechanism for partial breaking of the superconformal 
symmetry; the corresponding Goldstone field for broken 
scale and $S$-supersymmetry transformations should be 
an improved $\cN=1$ tensor supermultiplet \cite{DR}.
It would be interesting to re-derive our model
in the framework of superembeddings; see
\cite{Sor} for a review of the superembedding  approach. 
The construction presented in this paper should extend naturally 
to supersymmetric theories with $\cN >1$. In particular, 
in the case $\cN=2$, the Goldstone superfield is expected to be
the abelian $\cN=2$ vector supermultiplet described by a chiral constrained
superfield $W$ \cite{GSW}. Since $W$ contains the field strength 
$F_{mn}$ as one of its components, the resulting action should be 
of Born-Infeld type. 

\vskip.5cm

\noindent
{\bf Acknowledgements.}
Discussions with Arkady Tseytlin are gratefully acknowledged.
We thank Dima Sorokin for helpful comments on the first 
version of the manuscript.

\end{document}